# Progress in patterned wax stamp for prototyping of paper-based microfluidic analytical devices via injection molding

Zhizhi Zhou[a], Jiahuan Jiang[d], Yuanyuan Sun[a], Qing Qin[a], Sitong Yuan[a], Xilin Wang[a], Jianhua Jiang[a], Yifeng, Su[a], Xing Hu[a], Mingying Liu[b, *], Feng Yang[c, *]

[a] Key Laboratory of Research and Utilization of Ethnomedicinal Plant Resource of Hunan Province, Hunan Provincial Higher Education Key Laboratory of Intensive Processing Research on Mountain Ecological Food, College of Biological and Food Engineering, Huaihua University, Huaihua 418008, China

[b] School of Comprehensive Health Management, Xihua University, Chengdu, Sichuan 610039, China

[c] Key Laboratory of Magnetic Suspension Technology and Maglev Vehicle (Ministry of education of China), School of Electrical Engineering, Southwest Jiaotong University, Chengdu, Sichuan 610031, China

[d] Key Laboratory of Biorheological Science and Technology, Ministry of Education, Bioengineering College of Chongqing University, Chongqing 400044, China

*Corresponding should be addressed to Mingying Liu: kakarye@163.com and Feng Yang: yf@swjtu.edu.cn

## Abstract

In this study, we successfully developed two-dimensional (2D) paper-based microfluidic analytical devices (µPADs) using a hybrid technique of injection molding and embossing. This innovative approach involves passive or active delivery of molten wax onto a glass substrate through a sealed PDMS microfluidic chip, facilitating wax stamp creation post-demolding. Subsequent heating of the glass-stamping-paper assembly allows the molten wax to penetrate into the paper's cross-section, establishing distinct hydrophobic-hydrophilic patterns. The dimensions of these patterns can be adjusted with considerable flexibility.

We explored how the pattern size correlates with the melting duration and temperature. The method enables the integration of biological and chemical sensors into the patterns, thereby facilitating the production of complete µPADs. Our findings underscore this combined approach to manufacture large-scale, low-cost paper-based microfluidic sensors, particularly suitable for widespread use in resource-constrained settings.

## Introduction

Microfluidics is recognized as a dominant technology for rapid testing across various applications including point-of-care testing (POCT), food and drug safety testing, and environmental monitoring(Cate, Adkins et al. 2015, Noviana, Ozer et al. 2021, Brazaca, Imamura et al. 2024). Compared to conventional materials such as glass, polydimethylsiloxane (PDMS), Polymethyl Methacrylate (PMMA), etc., paper-based microfluidic analytical devices (µPADs) are more accessible and cost-effective. These attributes make µPADs particularly suitable for use in resource-limited settings or for applications requiring low-cost solutions. Additionally, it offers several advantages over conventional PDMS-based prototype devices produced via soft lithography, such as simplicity, biodegradability and the elimination of the need for external pumps or other sophisticated equipment(Yetisen, Akram et al. 2013, Xia, Si et al. 2016).

Numerous methods developed for constructing two-dimensional µPADs with hydrophilic and hydrophobic patterns. These include photolithography(Martinez, Phillips et al. 2007, He, Ma et al. 2013, Yu and Shi 2015), various printing techniques such as inkjet printing and wax printing (Maejima, Tomikawa et al. 2013, Zhang, Zhou et al. 2014, Yamada, Henares et al. 2015), wax screen-printing(Dungchai, Chailapakul et al. 2011), plasma etching(Li, Tian et al. 2008, Raj, Breedveld et al. 2019), cutting(Fenton, Mascarenas et al. 2009, Fu, Lutz et al. 2010, Chitnis, Ding et al. 2011, Xie, Zi et al. 2019), plotting(Nie, Zhang et al. 2012, Amin, Ghaderinezhad et al. 2017), writing and hand drawing(Cai, Wu et al. 2013), stamping(Cheng, Mazzeo et al. 2010, Curto, Lopez-Ruiz et al. 2013, Garcia, Cardoso et al. 2014, Mathaweesansurn, Thongrod et al. 2020), embossing(Thuo, Martinez et al. 2014, Juang, Chen et al. 2019), and others techniques(Gao, Liu et al. 2016). Each technique has limitations, particularly in efficient prototyping for low-cost applications. For example, wax and inkjet printing are preferred for µPADs fabrication due to their low cost, ease of use, and rapid production times. However, these methods often suffer from blurred boundaries and may lead to issues such as inkjet head clogging. Contemporary methods like stamping, embossing, plotting, and calendaring offer practical approaches to µPADs fabrication, although preparing molds for stamping can be both time-consuming and costly. While writing and hand drawing provide flexibility in deigned patterns for laboratory applications, scaling these methods for mass production in resource-limited settings remains challenging.

In this technical note, we introduce an innovative method for fabricating two-dimensional (2D) µPADs using injection wax molding and contact embossing. This technique involves the transfer of a wax stamp from a PDMS microchannel to filter paper via contact embossing. The molten wax serves as a barrier, enabling liquid, transferred into the sealed PDMS microfluidic channel where the wax stamp is created through passive pumping. Unlike the previously mentioned embossing methods, this approach allows for the creation of µPADs with hydrophobic and hydrophilic microstructures within few minutes, making it highly efficient for batch production. Furthermore, colorimetric assays were performed to evaluate the qualitative capabilities of these devices highlighting their potential advantages.

## Experimental section

### Materials and reagents

A solution of 2% ninhydrin was prepared by mixing 1.0g of ninhydrin with 40mg of $SnCl_2$ in 25 mL of deionized water. Polydimethylsiloxane (Sygard 184 silicone Elastomer Kit) was purchased from Dow Corning

(Midland, MI, USA). Green tea of Gougunao (Jiangxi, China) was used as the sample in this study. Laboratory filter paper was obtained from Beimujiang Company (Hangzhou, China). $58^{\#}$ Semi-refined paraffin wax was obtained from Petrochina Company (Liaoning, China). All reagents were used as received, without further purification. Ultrapure water with a resistivity of 18.0 MΩ•cm was used throughout the experiments.

**Preparation of total amino acid in tea leaves extract**

Two 3.0g samples of green tea were precisely weighed and then they were immersed into 300mL of water in two separate beakers. One sample was steeped in hot water (from 100 °C to 35 °C) for 45min and the other one was boiled at 100°C for the same duration to extract total amino acids for analysis. After cooling the solutions to room temperature, the filtrates were diluted to 500 mL using deionized water.

**Choice of paper and heating setting**

We selected laboratory filter paper as the substrate for paper-based microfluidic devices due to its cost-effectiveness, reproducibility, and biocompatibility. It is available in circle sheets' dimension of 30mm, that can be convenient for fabrication and embossing.

A digital heating plate, characterized by a smooth and uniform surface, was employed for the embossing step. Additional heating devices, including ovens and heat pipette guns, were utilized for curing PDMS and transferring wax, respectively.

**Facile Prototyping of 2D paper-based microfluidic devices**

The fabrication process for prototyping 2D patterns on paper-based microfluidic devices using a wax stamp and injection molding consists of the following two steps:

**Step 1: injection molding for patterning wax stamp**

This technique utilized an economical positive master with self-adhesive tape, which is patterned by laser engraving for PDMS replica molding. Subsequently, it facilitates passive pumping of melted wax into a PDMS microfluidic device to create a wax stamp formation. This stamp was heated and embossed onto a paper-based microfluidic device as depicted in Figure. 1. The process comprises: (1) designing the pattern using Auto Cad software; (2) creating the positive master pattern with a laser engraving machine; (3) casting the positive master pre-cured PDMS; (4) reversibly sealing the PDMS slab with a glass plate; (5) introducing the melted wax into the microfluidic channels; (6) cooling the PDMS device to solidify and form the wax stamp; (7) detaching the PDMS slab to reveal the solid wax (8) heating the wax stamp to enable penetration through the cross-section of the filter paper.

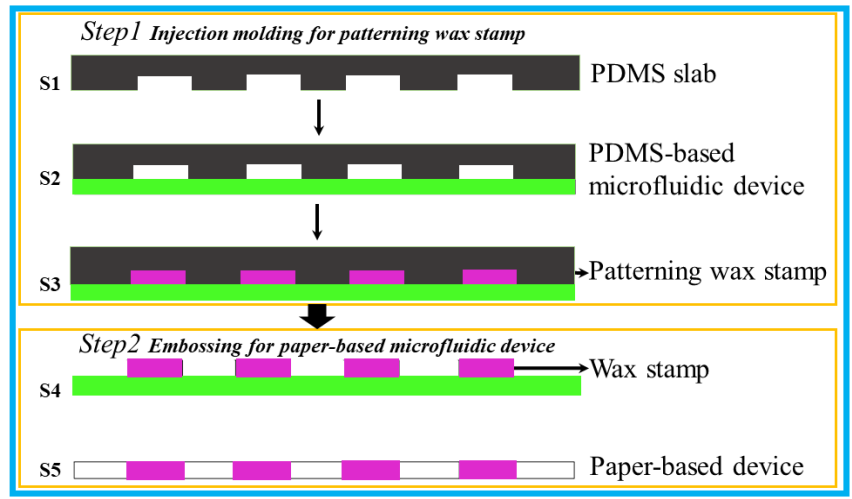

**Figure 1** The schematic of injection molding for fabrication of paper-based microfluidic device

**Step 2: Fabrication of paper-based microfluidic devices.**

The average pore size of the filter used in this study ranged 20~25 µm. During the spreading step, the wax from the stamp on the filter paper's surface melted, penetrating the filter paper both horizontally and vertically. To evaluate the extent of wax spreading, the heating temperature and time were adjusted from 100-130°C in increments of 10°C and from 0 to 30 seconds, respectively. These conditions resulted in hydrophobic barrier widths from 300 to 1300µm. Following the application of red ink to the patterned paper, the width of the hydrophobic barriers and hydrophilic zones were measured by using Image J software. To access the resolution of the proposed approach, the widths of the injected wax molds were compared with the final widths of the paper-based microfluidic devices.

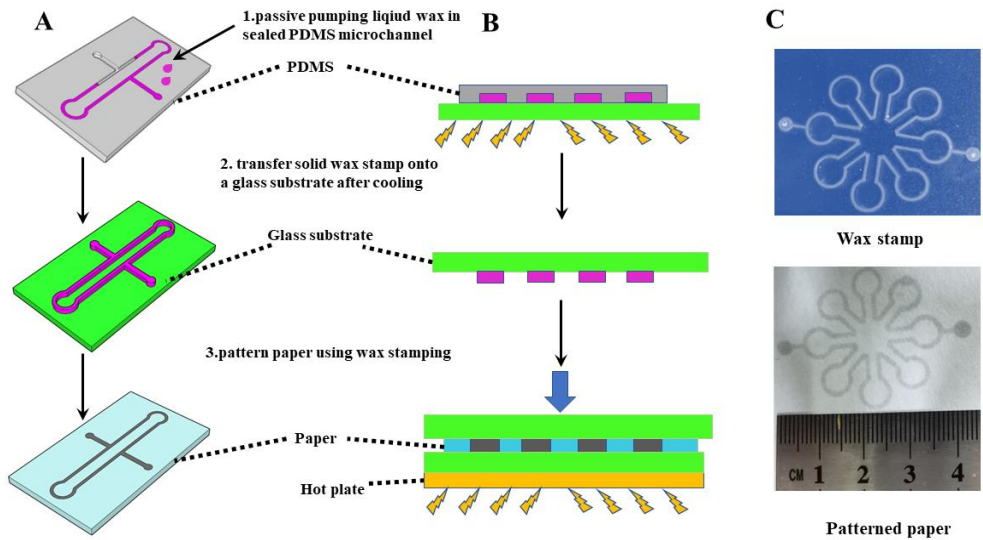

**Figure 2.** Prototype paper-based microfluidic devices fabricated by using injection wax molding and embossing technique. (A) oblique view, (B) cross section and (C) prototyping of wax stamp and patterned paper.

## Results and discussion

### Patterning wax stamp by injection molding

In this study, Auto CAD software was employed to design the pattern of the positive master for PDMS-based microfluidic devices. The positive master, equipped with self-adhesive tape, was prepared flexibly using laser engraving, a method referred to as facile and low-cost method(F-L) that was developed in our previous work(Zhou, Chen et al. 2017). Pre-cured PDMS was poured onto the patterned master, allowed to solidify, and removed. Subsequently, the punctured PDMS slab was cleaned and reversibly sealed onto a glass substrate to complete PDMS-based microfluidic device.

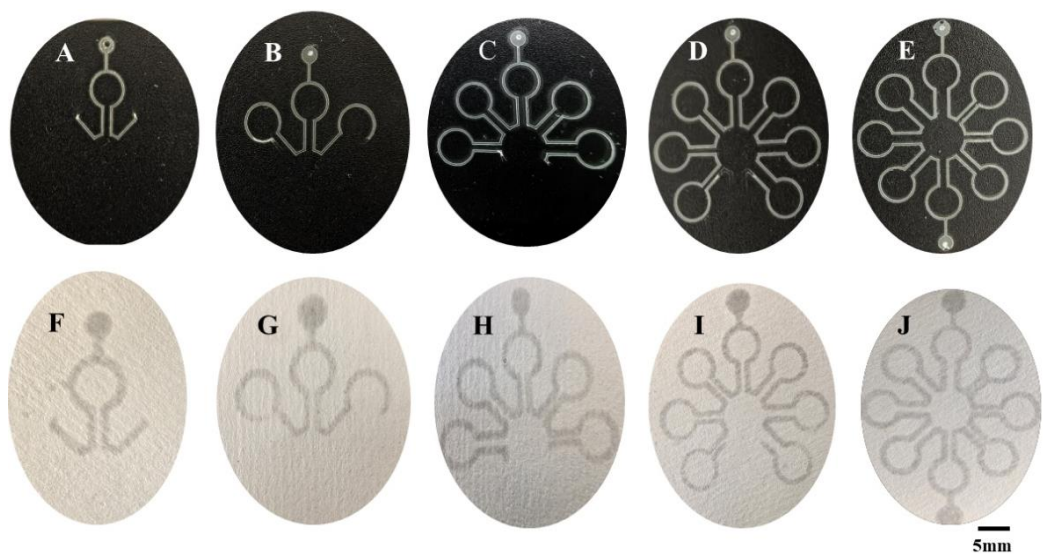

Figure 3. The photographs showing examples of wax stamps and paper-based devices with varies temperature and time by using a passive pumping method. Among them, μPAD (F) was fabricated by stamp (A), μPAD (G) was fabricated by stamp (B), μPAD (H) was fabricated by stamp (C), μPAD (I) was fabricated by stamp (D); μPAD (J) was fabricated by stamp (E).

Under wax melting conditions, it was observed for the first time that liquid wax at the inlet was drawn into the PDMS microfluidic channel by capillary action to form a patterned wax stamp (Figure 2A). The cross-sectional view of the prototype device is depicted in Figure 2B. The video detailing the wax stamp patterning process is shown in Supplementary Figure S1. The extent of wax streaming was enhanced by passive pumping with the stamp formation varying with temperature and time (Figure 3). The distance of wax injection was found to be correlated with the duration of passive injection, as illustrated in Figure 4, whereas the injection time inversely corelated with the temperature. Notably, as shown in Figure 4, the length of the wax increased from 0 to 111 mm when the duration increased from 1 to 190 seconds at 90°C. Subsequently, the minimum injection time was reduced to 85 seconds at 180°C, which the viscosity of the melting wax was decreased to shorten the injection time while the temperature increased. This phenomenon is highly analogous to findings reported by Beebe's group(Walker and Beebe 2002, Berthier and Beebe 2007). Moreover, no melted wax to be placed at the outlet to form a larger drop for passive driven, that is different from Beebe's work above. Beyond passive pumping, active injection can be expedited by vacuum assistance in several seconds is shown in Supplementary Figure S2.

**Table 1 Comparison of Injection molding and Existing conventional methods**

| Fabrication methods | Equipment | Cost | Batch fabrication | Resolution | Specialized skills | References |
|---|---|---|---|---|---|---|
| Photolithography | Photolithography machine | Very high | Medium | High | Yes | (Martinez, Phillips et al. 2007) |
| Printing | Customized printer | Very high | Yes | High | Yes | (Maejima, Tomikawa et al. 2013) |
| Etching | Plasma machine | Very high | Yes | Medium | Yes | (Raj, Breedveld et al. 2019) |
| Plotting | Plotter | Medium | Yes | Low | Yes | (Nie, Zhang et al. 2012) |
| Cutting | Cutter | High | Yes | High | Yes | (Thuo, Martinez et al. 2014, Xie, Zi et al. 2019) |
| Writing | Mark pen | Low | No | Low | No | (Cai, Wu et al. 2013) |
| Embossing | Computer numerical control machine (CNC) | High | Yes | High | Yes | (Juang, Chen et al. 2019) |
| Injection molding | Laser engraving machine | Very low | Yes | Very high | No | This work |

Compared to existing methods for producing wax patterning (Table 1), our proposed method offers enhanced flexibility in controlling the width of hydrophobic and hydrophilic patterns. Additionally, it does not require extensive expertise, sophisticated equipment, or significant resource expenditure. These advantages make it particularly suitable for implementation in resource-limited environment and low-cost diagnostic applications.

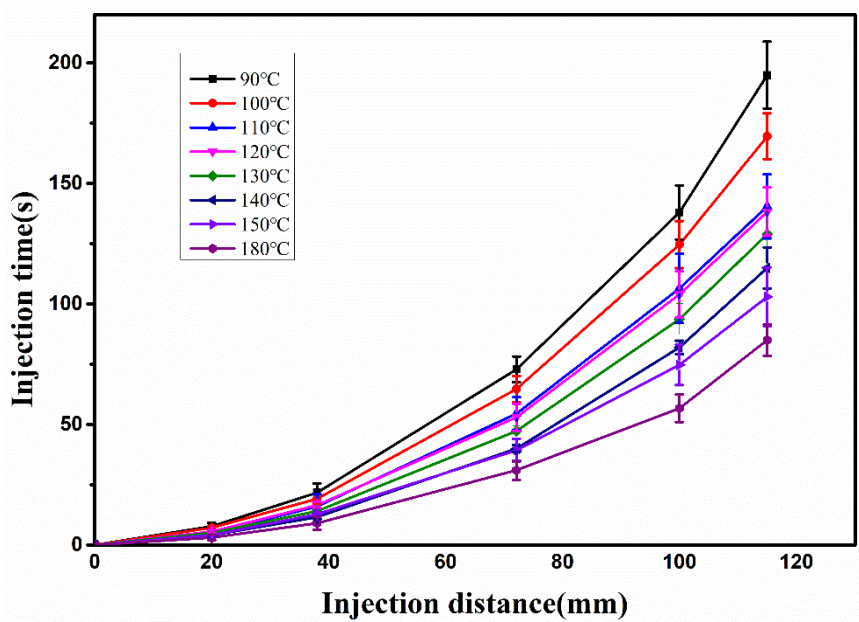

Figure 4. Plot of the distance of the molten wax stream as a function of the duration and temperature

**Characterization of hydrophobic-hydrophilic patterns in paper**

Figure 5(A) reveals several shapes pattern of wax stamp using injection method, and red ink was introduced to show the stamped hydrophilic area on the μPAD device is depicted in Figure 5(B).

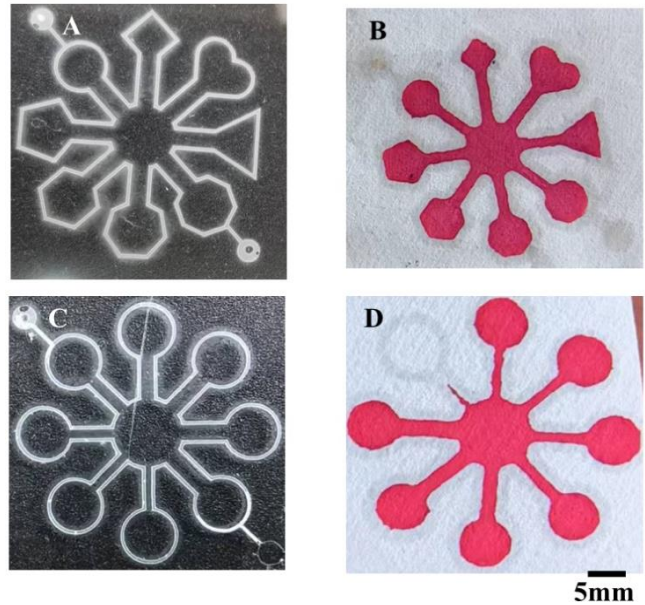

Figure. 5 (A) Several shapes of wax stamp for prototyping of µPAD device; (B)Wax stamped microfluidic pattern on filter paper with borders characterized by the hydrophilic-hydrophobic contrast (wetted with red ink); (C) Variation widths of the hydrophobic design by using injection molding (D) Variation of hydrophilic channel width with red color dependent on wax stamping

At it can be seen in Figure 6, under the condition of 3 seconds and 120 °C, the widths of the hydrophilic channels from 600 to 2000µm were investigated in the increments of 200µm. However, more than 5 seconds of stamping time should be excluded while the most of hydrophilic channels could be blocked due to the excess of the molten wax. It is evident that the mark-pattern width of the back side is similar to that of the front side, suggesting that the melted wax successfully penetrated through the filter paper during the hot embossing process. Upon cooling, the wax retained in the filter paper creates a hydrophobic barrier, enabling the directed flow of aqueous solutions (Figure 3J). That is, the wax-free hydrophobic independent detection zones or channels were defined. A linear equation was formulated to optimize the heating time and temperature is shown in Figure 6.

The device consists of eight distribution channels, six circle zones with a diameter of 6 mm on the end of the channels and one circle zone with a diameter of 8 mm in the center (Figure 3J).

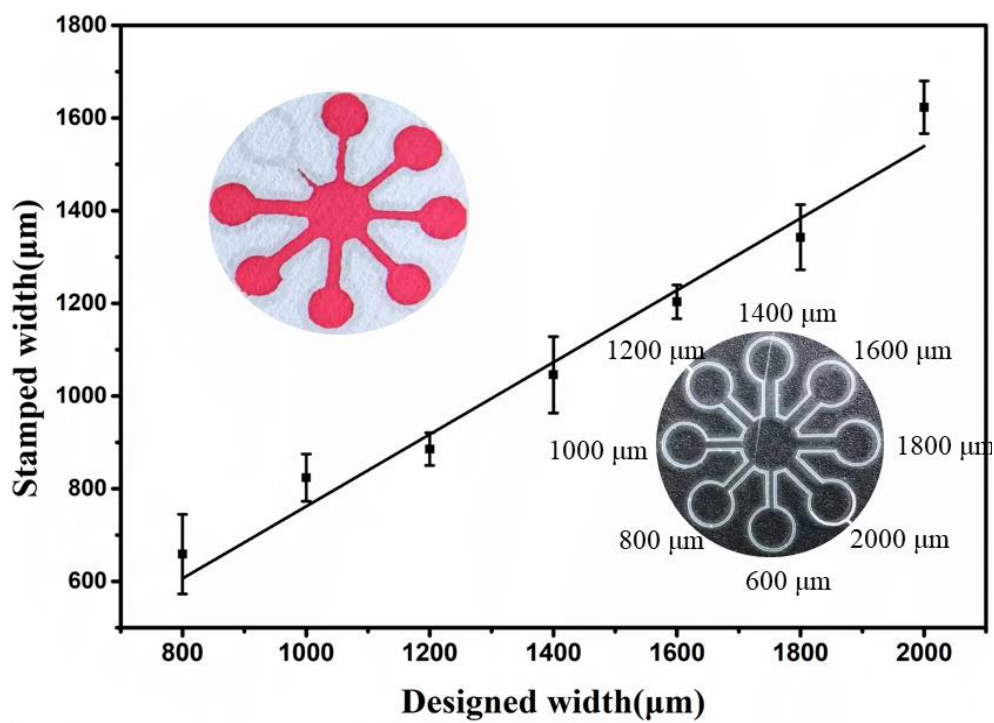

Figure.6 Variation of microfluidic channel width; the wax pattern fabricated is shown at bottom right of the figure; the hydrophilic pattern with red ink were shown at top left of the figure (error bar is from the standard deviation for 3 samples; the width of each channel was metered at 10 positions and the mean value was drawn as the red marked width),

linear curve fit $R^2$ is 0.975.

**Colorimetric assays**

In order to evaluate the effect of our proposal patterning method on the application, colorimetric detection of the total amino acid content of tea leaf extract as a model was carried out. The total amino acid content of tea leaf extract was provided in the section of experimental section mentioned above. 20μL of 2% ninhydrin solution was spotted in the center zone within 110 s, and then 1μL samples and standard agents were added in the detection zones of μPADs. Theses μPADs were transferred into oven at 65°C for 20min to generate a complete color reaction in the detection zones. The photo of the colorimetric assay was captured with a digital camera in dark room and kept in TIFF format (Figure 7A). The photo of the colorimetric assay was inverted by ImageJ software to obtain the mean gray values in testing zones. The results of the gray value were loaded into Origin software (version 9.0) to show a linear relationship between mean gray intensity (MGI) and standard amounts of amino acid concentrations (C: from 30 to 150 μg/mL) (Figure 4B). The formula between concentrations of glutamic acid and mean gray intensity is showed as followed.

$$MGI=0.16C\ (\mu g/ml)\ +31.8 \qquad (1)$$

a correlation coefficient of 0.9932 was obtained (Figure 7B). The concentration of the total amino acid content in the green tea leaves extract was caculated as 51.4 μg/ml and 43.3μg/ml in boiling water （sample 2） and hot water(sample 1) for 45min by using equation(1), respectively.

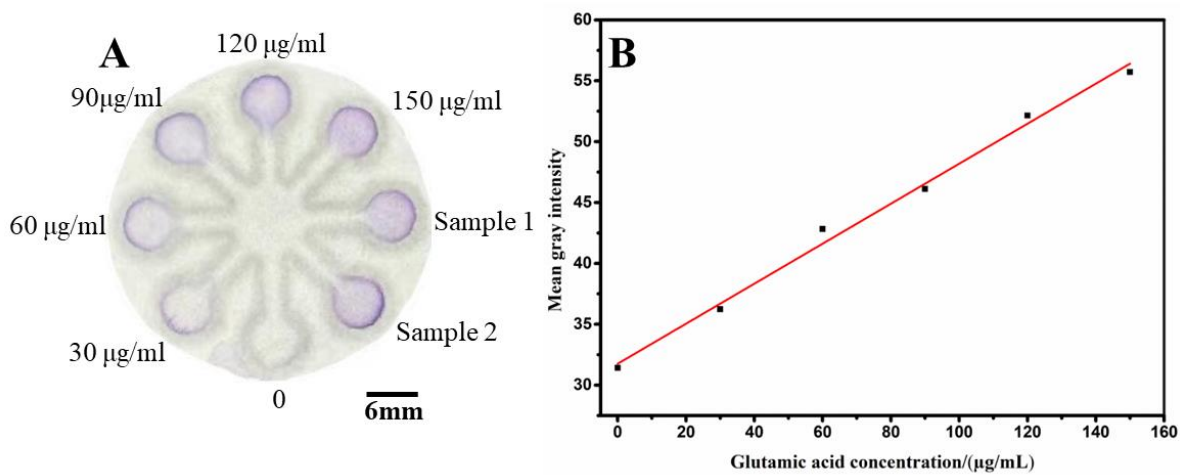

Figure 7. (A) Represent photo showing assay of glutamic acid and tea amino acid on paper-based device with different concentrations. (B) The linear correlation between the mean gray intensity and varies concentration of amino acids obtained from the photograph (A) (n=5).

**Conclusions**

In this paper, we introduce a passive injection molding method for fabrication of wax stamp. This technique is robust, convenient, inexpensive and suitable for resource-limited regions and countries. A linear relationship between the size of a hydrophobic pattern in paper and the size of a wax stamp was used to calculate. Finally, colorimetric assays have been successfully demonstrated the utility of this proposed method.

**Acknowledgement**

We acknowledge the project supported by the Research Foundation of Education Bureau of Hunan Province, China (Grant No. 23B0725), the Research Foundation of Huaihua university, China (Grant No. HHUY2021-16), the Foundation of Hunan Double First-rate Discipline Construction Projects (Grant No. SWGC-04) and the Project for Excellent Talents in Xihua University Grant (Z212031).